\documentclass[aps,superscriptaddress,showpacs,nofootinbib, twocolumn]{revtex4}
\usepackage{bm}
\usepackage{xcolor}
\usepackage{graphicx,color}
\usepackage{amsmath}
\usepackage{amssymb}
\usepackage{wasysym}

\newcommand{\be}{\begin{equation}}
\newcommand{\ee}{\end{equation}}
\newcommand{\bea}{\begin{eqnarray}}
\newcommand{\eea}{\end{eqnarray}}

\begin{document}

\title{Chaos and information in two dimensional turbulence}
 
\author{Daniel Clark}
\author{Lukas Tarra}%
 \author{Arjun Berera}
 \affiliation{%
 School of Physics and Astronomy, University of Edinburgh, JCMB,
\\King’s Buildings, Peter Guthrie Tait Road EH9 3FD, Edinburgh, United Kingdom.
}%

\date{\today}

\begin{abstract}
By performing a large number of fully resolved simulations of incompressible homogeneous and isotropic two dimensional turbulence, we study the scaling behavior of the maximal Lyapunov exponent, the Kolmogorov-Sinai entropy and attractor dimension. The scaling of the maximal Lyapunov exponent is found to be in good agreement with the dimensional predictions. For the cases of the Kolmogorov-Sinai entropy and attractor dimension, the simple dimensional predictions are found to be insufficient. A dependence on the system size and the forcing length scale is found, suggesting non-universal behavior. The applicability of these results to atmospheric predictability is also discussed.
\end{abstract}

\maketitle


\section{Introduction}

Turbulent fluid flows exhibit complex and, at first glance, apparently random motions. Consequently, our ability to predict their behavior is limited. Given that such fluids are governed by deterministic equations of motion, for example the Navier-Stokes equations for non-conducting fluids, their lack of exact predictability appears paradoxical. This aspect of turbulent flows can be understood as a consequence of deterministic chaos \cite{ott2002chaos, bohr2005dynamical} and an extreme sensitivity to initial conditions. As a result any error in measuring the state of the system, no matter how small, will be amplified as the system evolves, resulting in a finite predictability time. Turbulent flows are ubiquitous in the universe, occurring across a massive range of length scales and as such, quantifying their predictability may have wide reaching applications. Furthermore, fluid turbulence is in many ways representative of extended dynamical systems in general, and therefore such results may also be of more broad interest.

The study of the chaotic properties of dynamical systems began with the pioneering work of Lorenz \cite{lorenz1963deterministic}, exploring what is effectively a low dimensional model of the Navier-Stokes equations. These ideas were then employed by Ruelle and Takens \cite{ruelle1971nature} to describe a mechanism by which turbulence can be generated in a fluid flow. In applying the methods of chaos theory to the study of turbulence, we consider the properties of individual trajectories through a suitably defined state space of the system. This is in contrast to the more common approach of studying the statistical properties of turbulence through averaging \cite{batchelor1953theory} over time, space or numerous realizations of the system.

Starting with the pioneering studies of Leith and Kraichnan \cite{leith1971atmospheric, leithkraich}, a large body of work dedicated to the study of predictability in turbulent fluid flows has formed. These initial studies made use of turbulent closure models in order to render the problem computationally feasible. Unfortunately, these closure models have a number of shortcomings that may reduce their ability to correctly quantify predictability in turbulence. Perhaps most notably, as they are typically defined in terms of ensemble averaged quantities, they do not provide any information about the spatial structure of the flow which is likely to influence predictability. Additionally, while these models are well-known to give excellent agreement with the K41 theory of turbulence \cite{k41}, the validity, or lack thereof, of K41 itself is still unsettled \cite{frisch}. As such, the applicability of the results of these models to true fluid turbulence may be limited. 

As computing power increased, it became possible to perform predictability measurements in direct numerical simulations (DNS) of turbulence. Since such simulations fully resolve all the relevant scales of the system in both space and time, they provide far more reliable results when compared to closures. However, such simulations come with a large computational expense, so progress has been made at a moderate rate. This is especially true for predictability studies where the computational cost is at least twice that of a standard simulation. 

Historically, the majority of work regarding the predictability of fluid turbulence has centered around the measurement of the maximal Lyapunov exponent of the system. This gives a measure of the rate at which nearby trajectories in the state space diverge, and thus also provides a measure of the predictability time of the system. Due to the aforementioned computational expense, these studies began by focusing on the less demanding case of two dimensional turbulence \cite{kida90, boffetta01} as well as moderate Reynolds number three dimensional turbulence \cite{kida92}. More recently, a number of studies at higher Reynolds number in three dimensions have been performed \cite{berera2018chaotic, boffetta2017chaos, mohan17}, as well as a study into predictability in magnetohydrodynamic turbulence \cite{ho2019chaotic}. Each of these hydrodynamical studies independently found that the maximal Lyapunov exponent scaled faster with the Reynolds number than predicted by Ruelle \cite{ruelle1979microscopic} using the K41 theory. This is of particular interest, as by using the multi-fractal model \cite{benzi1984multifractal}, developed in an attempt to capture the effects of internal intermittency in turbulence, it can be shown that the maximal exponent should scale slower than predicted by Ruelle \cite{crisanti1993predictability}, opposite to what was found in DNS. 

It is also possible to study more than simply the behavior of the maximal Lyapunov exponent. There exist as many exponents as there are degrees of freedom in the system and these are said to form a Lyapunov spectrum. However, since the calculation of each additional exponent desired comes with further computational cost, the study of the Lyapunov spectrum in fluid turbulence is still at a comparatively early stage when compared to that of the maximal exponent, although seems to be following the same path of development. Initially, studies were restricted to shell models \cite{grappin1986computation, yamada1987lyapunov, yamada1988lyapunov} but soon progressed to DNS studies in two \cite{grappin1987computation, grappin1991lyapunov} and three dimensional Poiseuille flow \cite{keefe1992dimension}, as well as a highly symmetric homogeneous and isotropic turbulence (HIT) system \cite{van2006periodic}. By measuring all of the positive Lyapunov exponents of a system, the Kolmogorov-Sinai (KS) entropy, which quantifies the rate of information production of the system and gives a more accurate quantification of predictability, can be estimated. 

Such is the computational expense in measuring the KS entropy, that only very recently, and at moderate Reynolds number, has the scaling of the KS entropy in three dimensional HIT been measured \cite{berera2019info}. Here it was found that the entropy scaled slower with the Reynolds number than predicted using dimensional arguments and K41, although only marginally, however, the related attractor dimension was found to scale faster. Both results should be interpreted with some caution given the relatively low Reynolds numbers obtained. Owing to the reduced computational effort required to study two dimensional turbulence, a small investigation into the attractor dimension scaling was performed in \cite{grappin1987computation}. Computing power is now such that a systematic study of the chaotic properties of two dimensional turbulence can be performed and is the focus of this investigation. Features of two dimensional turbulence, although not truly realizable itself, can be seen across a wide range of fluid systems. For example, in systems where one dimension is constrained compared to the others, such that the fluid exists in a thin layer, two dimensional effects have been observed \cite{xia}. Moreover, in atmospheric measurements here on Earth \cite{nastrom} and elsewhere in the solar system \cite{young} evidence of two dimensional phenomenology has been found.  As such, understanding the predictability of two dimensional HIT may be of more relevance to atmospheric predictability than the three dimensional case in many situations.

This paper is organized as follows: in section \ref{2} we introduce a number of scaling predictions for the maximal Lyapunov exponent, attractor dimension and KS entropy in two dimensional HIT. Next, in section \ref{3} we discuss the numerical methods used in computing the chaotic properties we are interested in, focusing on the computation of the Lyapunov spectrum. We then present the results of our numerical study in section \ref{4} and compare them to the theoretical predictions. Here, we find unexpected corrections to the theoretical predictions which suggest an influence from the system size and forcing length scale, hinting at a lack of universality. Finally, in section \ref{5} we  discuss the implication of our results for two dimensional HIT as a whole as well as possible applications to less idealized fluid systems.

\section{Scaling predictions for two dimensional turbulence}\label{2}

For the case of three dimensional HIT, there exist a number of theoretical predictions for the scaling behavior of the maximal Lyapunov exponent, attractor dimension and KS entropy. The simplest of such predictions are all based on dimensional arguments and the K41 theory. These ideas can be applied in an analogous way to two dimensional HIT where the dual cascade picture \cite{kraich67, leith68, batch69}, caused by conservation of both energy and enstrophy, modifies the expected scaling behavior. 

The scaling behavior of the maximal Lyapunov exponent, $\lambda_1$, in two dimensional HIT can be found by following the Ruelle argument \cite{ruelle1979microscopic} that on dimensional grounds it should be proportional to the inverse of the fastest timescale in the flow. In three dimensional HIT this is the Kolmogorov timescale, corresponding to the small scales of the flow, which then suggests a scaling with the Reynolds number. On the other hand, for two dimensional HIT, assuming the energy spectrum is $E(k) \sim k^{-3}$ in the direct cascade and therefore neglecting, for now, any logarithmic corrections, there is only one timescale throughout the entire direct enstrophy cascade. This timescale, $\tau$ is determined solely by the enstrophy dissipation rate, $\eta$, and is given by $\tau \sim \eta^{-1/3}$. As such, we then have \begin{equation}\label{eq1}
\lambda_1 \sim \frac{1}{\tau} \sim \eta^{\frac{1}{3}}.
\end{equation} Therefore, at odds with the three dimensional case, $\lambda_1$ scales independently of the Reynolds number. Furthermore, this suggests that in some sense the small scales of the flow retain information about the larger scales.

Turning to the KS entropy, once again on dimensional grounds alone we can estimate the scaling behavior. In this instance, the entropy should scale with the fastest timescale in the flow multiplied by the total number of excited modes, see for example \cite{shaw, wang1992varepsilon} for a description of this argument applied to three dimensional turbulence. To apply this method to two dimensional flows, we need to determine the scaling of the total number of excited modes, which is also the scaling of the attractor dimension, $\mathrm{dim}(A)$. We do so by considering the ratio of the largest scales in the flow, $L$, to the smallest given by the dissipation length scale, $\chi = (\nu^3/\eta)^{1/6} $ where $\nu$ is the viscosity. This gives us \begin{equation}\label{eq2}
\mathrm{dim}(A) \sim \left(\frac{L}{\chi}\right)^2 \sim \mathrm{Re},
\end{equation} which in turn implies that the KS entropy, $h_{\mathrm{KS}}$, will scale as \begin{equation}\label{eq3}
h_{\mathrm{KS}} \sim \frac{1}{\tau} Re = \eta^{\frac{1}{3}} \mathrm{Re}.
\end{equation}

It has been shown by Kraichnan that in order for a constant enstrophy flux in the direct cascade of two dimensional turbulence to exist there must be a logarithmic correction to the energy spectrum \cite{kraich71}. This correction will then affect the predicted scaling behavior of the various chaotic quantities we are interested in. This was considered by Ohkitani \cite{ohkitani} and introduces an additional logarithmic dependence on the Reynolds number to each quantity as follows \begin{equation}\label{eq4}
\lambda_1 \sim \left(\eta \log \mathrm{Re} \right)^{\frac{1}{3}},
\end{equation} \begin{equation}\label{eq5}
\mathrm{dim}(A) \sim \mathrm{Re} \left( \log \mathrm{Re} \right)^{\frac{1}{3}},
\end{equation} and \begin{equation}\label{eq6}
h_{\mathrm{KS}} \sim \eta^{\frac{1}{3}} \mathrm{Re} \left( \log \mathrm{Re} \right)^{\frac{2}{3}}.
\end{equation} These differ from the previous predictions only by logarithmic factors and thus may be hard to distinguish in practice, as such we will not pursue these strongly here. 

Finally, in \cite{ruelle1982large} scaling predictions for the KS-entropy and attractor dimension were derived for three dimensional turbulence. These results were then extended to the two dimensional case by Lieb \cite{lieb}. If the energy dissipation rate, $\varepsilon$, is taken to be constant throughout the fluid for the KS entropy, we have \begin{equation}\label{eq7}
h_{\mathrm{KS}} \sim \frac{\varepsilon}{\nu^2}V,
\end{equation} where V is the volume of the system. For the attractor dimension it is found that \begin{equation}\label{eq8}
\mathrm{dim}(A) \sim \sqrt{\frac{\varepsilon}{\nu^3}} V.
\end{equation}

We note here that there also exist a number of more mathematically rigorous scaling laws for some of these quantities, see for example \cite{constan}. These are typically expressed in terms of a generalized Grashof number which can be related to the Reynolds number. The dimensional scaling laws in Eqs. \ref{eq1}-\ref{eq8} are consistent with the rigorous upper bounds in \cite{constan}, thus we will focus on these simpler dimensional predictions in this work.
\section{Numerical method}\label{3}

In order to complete a model-independent study of the chaotic properties of incompressible two-dimensional HIT, we perform DNS of the Navier-Stokes equations in two spatial dimensions with a large scale hypo-viscous dissipation term \begin{equation}
\begin{split}
\partial_t u_i + u_j \partial_j u_i = -\partial_i P &+ \nu \nabla^2 u_i + \mu \nabla^{-2} u_i + f_i, \\
\partial_i u_i = 0, &\quad i,j = 1, 2.
\end{split}
\end{equation} Here, $\bm{u}(\bm{x},t)$ is the velocity field, $P(\bm{x},t)$ is the pressure field, $\mu$ is the hypo-viscosity and $\bm{f}(\bm{x},t)$ is an external force that we will specify and discuss later in this section. To obtain our results we have made use of the EddyBurgh code \cite{EddyBurgh}, a modification of that described in \cite{yoffe2013investigation}, and as such we make use of the pseudospectral method with full dealiasing using the two-thirds rule. In order to ensure the flow is well resolved, we ensure that $k_{\mathrm{max}}/k_d \gtrsim 1.25$ for all our simulations, where $k_d = 1/l_d$. In \cite{keefe1992dimension} it was found that insufficient resolution led to an underestimation of the attractor dimension, though, by following the criteria above in \cite{berera2019info} we found such issues were avoided. In order to study the effect of the physical size of the domain on the chaotic properties of the system, we have performed simulations in periodic boxes of side lengths $\pi/2, \pi$ and $2\pi$. Details of all simulations performed can be found in Tables \ref{tab:my-table}, \ref{tab:my-table1} and \ref{tab:my-table2}.

To obtain a stationary state some form of large-scale dissipation is necessary, as otherwise the inverse cascade will eventually lead to the formation of a condensate on the scale of the system size \cite{kraich67}. In real world flows which show two dimensional behavior, the large scale dissipation is given by friction between the two dimensional flow and the three dimensional system it is contained within. As such, friction terms which depend on the fluid velocity either linearly or quadratically are often used \cite{boffettaecke}. However, such terms have an effect on all scales of the flow, and given that our predictions in Eqs. \ref{eq1}-\ref{eq6} depend on the small-scale enstrophy dissipation rate, we opt for a large-scale dissipation that effectively does not act on the small-scales. As a consequence of the large computational cost of our simulations, we have not tested how our results would be affected by the use of friction as opposed to the inverse Laplacian used here. However, any difference should be small. This inverse Laplacian term makes more sense in Fourier space where it becomes $k^{-2}$, which highlights that it most strongly influences the large length scales of the flow. To quantify the effects of this term, we use the hypo-viscous Reynolds number, Re$_{\mu}$, where \begin{equation}
\textrm{Re}_{\mu} = \frac{u}{\mu L^3}.
\end{equation} This term is derived from the ratio of inertial to hypo-viscous forces. As such, when it is small, the hypo-viscous term is dominant at the large scales. This allows us to ensure no large scale condensate has formed,

Notably, we do not employ any form of hyper-viscosity, which is very often used in two dimensional HIT simulations to increase the enstrophy inertial range. It is arguable that the use of hyper-viscosity is akin to that of an effective viscosity employed in methods such as large-eddy simulation, and as such acts as a form of closure. Hence, we choose to avoid any ambiguity stemming from the use of hyper-viscosity in this study.

Throughout this work, when we refer to the Reynolds number, Re, we are considering the integral-scale Reynolds number. To define this we need to first define the integral length scale, $L$, which gives the rough size of the largest eddys in the flow.  It can be shown \cite{DavidsonBook} by considering the two point second order longitudinal velocity correlation function, that in two dimensions $L$ is given by \begin{equation}
L = \frac{2}{E}\int_{0}^{\infty} dk \, E(k)k^{-1}. 
\end{equation} The integral scale Reynolds number is then given by \begin{equation}
\mathrm{Re} = \frac{UL}{\nu},
\end{equation} where $U$ is the RMS velocity.

\begingroup
\squeezetable
\begin{table}[!ht]
\resizebox{0.5\textwidth}{!}{%
\begin{tabular}{lllllllllllll}
\hline
$h_{\mathrm{KS}}$ & $\sigma$ & $N_s$ & $N_e$ & dim$(A)$ & $\eta$ & Re & $\nu$ & $k_f$  & $k_{\mathrm{max}}$ & $\chi$ & Re$_{\mu} $\\ \hline
0.97  & 0.14 & 1037 & 19  & 49.37  & 0.028 & 183  & 0.001    & 3  & 20  & 0.057 & 1  \\
2.88  & 0.22 & 1042 & 35  & 75.33  & 0.196 & 39   & 0.003    & 5  & 20  & 0.072 & 3  \\
1.14  & 0.11 & 1040 & -  & 67.62  & 0.474 & 10   & 0.005    & 7  & 20  & 0.080 & 7  \\
0.74  & 0.09 & 280  & 23  & 46.23  & 0.016 & 210  & 0.0008   & 3  & 20  & 0.056 & 1  \\
1.23  & 0.13 & 513  & 28  & 62.63  & 0.023 & 380  & 0.0005   & 3  & 41  & 0.042 & 1  \\
1.42  & 0.18 & 1900 & 24  & 57.58  & 0.029 & 281  & 0.001    & 3  & 41  & 0.057 & 3  \\
1.82  & 0.25 & 1900 & 26  & 62.24  & 0.058 & 400  & 0.001    & 3  & 41  & 0.051 & 4  \\
2.51  & 0.31 & 1900 & 30  & 72.98  & 0.090 & 492  & 0.001    & 3   & 41  & 0.047 & 4  \\
2.21  & 0.31 & 1900 & 28  & 69.07  & 0.113 & 560  & 0.001    & 3  & 41  & 0.045 & 6  \\
2.40  & 0.34 & 1900 & 31  & 74.34  & 0.137 & 632  & 0.001    & 3   & 41  & 0.044 & 6  \\
2.66  & 0.38 & 1900 & 32  & 77.06  & 0.168 & 694  & 0.001    & 3   & 41  & 0.043 & 8  \\
1.77  & 0.29 & 423 & 13  & 29.57  & 0.564 & 64  & 0.0085    & 3   & 41  & 0.101 & 2  \\
4.03  & 0.30 & 1900 & 51  & 114.58 & 0.106 & 204  & 0.001    & 5   & 41  & 0.046 & 9  \\
5.45  & 0.46 & 1900 & 56  & 128.30 & 0.213 & 298  & 0.001    & 5   & 41  & 0.041 & 11 \\
6.01  & 0.55 & 1900 & 59  & 135.38 & 0.319 & 358  & 0.001    & 5   & 41  & 0.038 & 14 \\
6.36  & 0.58 & 1900 & 60  & 139.35 & 0.424 & 406  & 0.001    & 5   & 41  & 0.036 & 19 \\
7.21  & 0.68 & 1900 & 65  & 149.38 & 0.525 & 456  & 0.001    & 5   & 41  & 0.035 & 20 \\
7.87  & 0.74 & 1900 & 68  & 158.04 & 0.635 & 496  & 0.001    & 5   & 41  & 0.034 & 23 \\
8.07  & 0.35 & 1900 & 74  & 168.57 & 0.229 & 180  & 0.001    & 7   & 41  & 0.040 & 9  \\
12.17 & 0.50 & 1900 & 84  & 199.31 & 0.458 & 276  & 0.001    & 7   & 41  & 0.036 & 9  \\
14.94 & 0.61 & 1900 & 94  & 226.20 & 0.692 & 339  & 0.001    & 7   & 41  & 0.034 & 11 \\
17.65 & 0.71 & 1900 & 99  & 250.44 & 0.910 & 403  & 0.001    & 7   & 41  & 0.032 & 11 \\
20.06 & 0.79 & 1900 & 108 & 277.12 & 1.153 & 459  & 0.001    & 7   & 41  & 0.031 & 11 \\
22.39 & 0.87 & 1900 & 109 & 302.59 & 1.381 & 501  & 0.001    & 7   & 41  & 0.030 & 13 \\
16.02 & 0.51 & 1900 & 111 & 257.03 & 0.556 & 182  & 0.001    & 9   & 41  & 0.035 & 12 \\
23.57 & 0.67 & 1900 & 129 & 320.14 & 1.100 & 296  & 0.001    & 9   & 41  & 0.031 & 10 \\
18.07 & 0.41 & 1900 & 139 & 304.91 & 0.592 & 112  & 0.001    & 11  & 41  & 0.035 & 15 \\
27.69 & 0.59 & 1900 & 157 & 375.80 & 1.184 & 191  & 0.001    & 11  & 41  & 0.031 & 14 \\
1.38  & 0.17 & 1557 & 27  & 57.59  & 0.029 & 279  & 0.001    & 3   & 84  & 0.057 & 3  \\
4.70  & 0.40 & 1123 & 53  & 122.08 & 0.210 & 281  & 0.001    & 5   & 84  & 0.041 & 15 \\
4.10  & 0.30 & 1195 & 50  & 113.11 & 0.106 & 207  & 0.001    & 5   & 84  & 0.046 & 9  \\
1.55  & 0.22 & 254  & 16  & 32.78  & 0.263 & 86   & 0.005    & 3   & 41  & 0.088 & 2  \\
3.58  & 0.32 & 744  & 31  & 66.31  & 0.727 & 47   & 0.005    & 5   & 41  & 0.075 & 5  \\
4.26  & 0.29 & 671  & -  & 96.77  & 1.012 & 25   & 0.0045   & 7   & 41  & 0.067 & 11 \\
2.32  & 0.26 & 332  & 21  & 51.18  & 0.220 & 252  & 0.002    & 3   & 41  & 0.058 & 2  \\
9.08  & 0.35 & 164  & 82  & 211.24 & 0.144 & 433  & 0.0004   & 5   & 84  & 0.028 & 4  \\
1.32  & 0.25 & 526  & 26  & 69.18  & 0.022 & 436  & 0.00044  & 3   & 41  & 0.040 & 1  \\
14.29 & 0.74 & 446  & 120 & -   & 0.417 & 999  & 0.0003   & 5  & 84  & 0.020 & 8  \\
18.08 & 0.55 & 450  & 131 & -   & 0.385 & 397  & 0.0004   & 7  & 84  & 0.023 & 6  \\
24.87 & 1.05 & 418  & 177 & -   & 0.696 & 1297 & 0.0003   & 5  & 84  & 0.018 & 9  \\
29.89 & 0.69 & 399  & 179 & -   & 0.698 & 354  & 0.0004   & 9  & 84  & 0.021 & 7  \\
23.64 & 0.56 & 464  & 150 & -   & 0.710 & 218  & 0.0006   & 9  & 84  & 0.026 & 6  \\
14.89 & 0.49 & 125  & 189 & 470.54 & 0.110 & 2187 & 0.000085 & 5  & 169 & 0.013 & 7  \\
4.18  & 0.19 & 132  & 89  & 218.96 & 0.018 & 2631 & 0.000075 & 3  & 169 & 0.017 & 2  \\
36.01 & 0.91 & 212  & 334 & 848.38 & 0.320 & 2000 & 0.000085 & 7  & 169 & 0.011 & 8  \\
20.86 & 0.67 & 130  & 232 & -   & 0.146 & 5902 & 0.000075 & 3  & 169 & 0.012 & 2  \\
22.46 & 0.51 & 145  & 127 & 313.20 & 2.486 & 58   & 0.002    & 11 & 41  & 0.038 & 16 \\
2.99  & 0.26 & 577  & 75  & - & 2.770 & 9    & 0.004    & 11 & 41  & 0.053 & 56  \\ \hline
\end{tabular}
}
\caption{Simulation parameters for $k_{\textrm{min}} = 1$ data: $\sigma$ is the standard deviation of the Kolmogorov-Sinai entropy, $N_s$ is the number of samples used in determining the entropy and attractor dimension, $N_e$ is the number of positive Lyapunov exponents and $\chi = (\nu^3/\eta)^{1/6}$ is the two dimensional analogue of the Komogorov length scale. For the attractor dimension a `-' indicates insufficient exponents were obtained to compute the dimension.}
\label{tab:my-table}
\end{table}
\endgroup

\begingroup
\squeezetable
\begin{table}[!ht]
\resizebox{0.5\textwidth}{!}{%
\begin{tabular}{lllllllllllll}
\hline
$h_{\mathrm{KS}}$ & $\sigma$ & $N_s$ & $N_e$ & dim$(A)$ & $\eta$ & Re & $\nu$ & $k_f$  & $k_{\mathrm{max}}$ & $\chi$ & Re$_{\mu} $\\ \hline
0.89  & 0.20 & 886 & 11  & 23.78  & 0.084 & 217  & 0.001 & 4  & 40 & 0.048 & 2  \\
4.93  & 0.47 & 231 & 61  & 161.52 & 0.066 & 3323 & 0.0001 & 4 & 83 & 0.016 & 3  \\
14.01 & 1.05 & 232 & 141 & -   & 0.252 & 2856 & 0.0001 & 6 & 83 & 0.013 & 9  \\
3.69  & 0.51 & 232 & 33  & 83.73  & 0.322 & 771  & 0.0003 & 6 & 83 & 0.021 & 13 \\
2.25  & 0.34 & 122 & 28  & 67.96  & 0.078 & 1284 & 0.0002 & 4 & 83 & 0.022 & 5  \\
1.10  & 0.25 & 240 & 9   & 20.78  & 0.587 & 164  & 0.002 & 4  & 83 & 0.049 & 3  \\
1.49  & 0.28 & 244 & 13  & 30.04  & 0.232 & 266  & 0.001 & 4  & 83 & 0.040 & 4  \\
0.74  & 0.24 & 443 & 6  & 12.59  & 0.770 & 59  & 0.005 & 4  & 83 & 0.074 & 3  \\
3.29  & 0.44 & 521 & 20  & 47.12  & 0.86 & 276  & 0.001 & 6  & 83 & 0.032 & 5  \\
3.25  & 0.52 & 578 & 15  & 38.67  & 1.74 & 164  & 0.002 & 6  & 83 & 0.041 & 5  \\ \hline
\end{tabular}
}
\caption{Simulation parameters for $k_{\textrm{min}} = 2$ data.}
\label{tab:my-table1}
\end{table}
\endgroup

\begingroup
\squeezetable
\begin{table}[!ht]
\resizebox{0.5\textwidth}{!}{%
\begin{tabular}{lllllllllllll}
\hline
$h_{\mathrm{KS}}$ & $\sigma$ & $N_s$ & $N_e$ & dim$(A)$ & $\eta$ & Re & $\nu$ & $k_f$  & $k_{\mathrm{max}}$ & $\chi$ & Re$_{\mu} $\\ \hline
0.89 & 0.41 & 123 & 6  & 15.84 & 1.436 & 321  & 0.0006 & 8  & 167 & 0.023 & 34 \\
0.92 & 0.37 & 305 & 8  & 20.45 & 0.484 & 300  & 0.0004 & 8  & 339 & 0.023 & 28 \\
1.71 & 0.59 & 558 & 10 & 24    & 3.675 & 421  & 0.00075 & 8 & 167 & 0.022 & 35 \\
1.23 & 0.39 & 590 & 15 & 26.6  & 2.148 & 635  & 0.00045 & 8 & 167 & 0.019 & 37 \\
1.34 & 0.33 & 120 & 16 & 38.05 & 0.056 & 1653 & 0.00005 & 8 & 339 & 0.011 & 22 \\
0.22 & 0.13 & 392 & 5  & 8.11  & 0.619 & 9    & 0.005 & 8   & 83  & 0.077 & 7  \\
4.01 & 0.73 & 588 & 31 & 68.85 & 0.893 & 778  & 0.00015 & 12 & 167 & 0.012 & 57 \\
3.93 & 0.45 & 118 & 28 & 65    & 0.177 & 1092 & 0.0001 & 12  & 167 & 0.013 & 9  \\
1.68 & 0.33 & 210 & 11 & 29.15 & 0.546 & 256  & 0.0004 & 12  & 167 & 0.022 & 8  \\
1.32 & 0.26 & 300 & 6  & 18.06 & 0.324 & 137  & 0.00075 & 8 & 167 & 0.033 & 7  \\
1.9  & 0.52 & 580 & 9  & 23.52 & 0.438 & 336  & 0.0005 & 8  & 167 & 0.026 & 15 \\ \hline
\end{tabular}
}
\caption{Simulation parameters for $k_{\textrm{min}} = 4$ data.}
\label{tab:my-table2}
\end{table}
\endgroup

\subsection{Forcing}

Due to the presence of dissipative terms in the Navier-Stokes equations, energy must be injected into the system in order for a statistically stationary state to be achieved. To ensure our results are independent of the way this energy is injected, we have made use of two different forcing functions. Indeed, we find the choice of forcing does not affect our results. The first of these functions is defined as \begin{equation}
\boldsymbol{f}(\boldsymbol{k},t) =\begin{cases}
 (\varepsilon/2E_{f})\boldsymbol{u}(\boldsymbol{k},t) \quad &\text{if } |\bm{k}|\approx k_f, \\
 0 \qquad &\text{else,}
\end{cases}
\end{equation} where $E_f = E(k_f)$ is the energy in the forcing band and $\varepsilon$ is the energy injection rate. More explicitly, the forcing acts on the ring of modes satisfying $k_f - 1/2 < |\bm{k}| \leq k_f + 1/2$. This forcing has been widely used for studies of three dimensional turbulence and allows the rate of energy injection to be held constant in time. 

The second forcing employed is a delta-correlated in time stochastic force with amplitude \begin{equation}
f_{\mathrm{amp}} = \begin{cases} \sqrt{\frac{2\epsilon}{dt}} \quad &\text{if } |\bm{k}|\approx k_f, \\
 0 \qquad &\text{else,}
\end{cases} 
\end{equation} where $dt$ is the simulation time step. This choice then ensures that $\langle \bm{u} \cdot \bm{f} \rangle = \varepsilon$, i.e. on average the energy injection is given by $\varepsilon$. This method of forcing is used commonly in simulations of two dimensional turbulence. Once again the forcing function is active only on modes which satisfy $k_f - 1/2 < |\bm{k}| \leq k_f + 1/2$. 

\subsection{Computation of the Lyapunov spectrum}

In chaotic systems there exists a Lyapunov exponent for every degree of freedom. As such, it is possible to define a set of such exponents, arranged in descending order, known as the Lyapunov spectrum. This concept can be defined in more formal terms and a good account of this for the case of fluid turbulence is given in \cite{ruelle1982large}. As is typical in the literature \cite{ott2002chaos}, we take the KS entropy to be given  by the sum of positive Lyapunov exponents \begin{equation}
h_{KS} = \sum_{\lambda_i > 0} \lambda_i.
\end{equation} Therefore, in each case we need to measure a number of exponents. Unfortunately, the method to compute these exponents comes with a number of computational challenges. Firstly, $\textit{a priori}$ we do not know in advance the number of positive exponents. Secondly, each exponent requires the simultaneous integration of another velocity field, and finally, many iterations are required to obtain averaged values for the exponents. Consequently, computing the KS entropy for fully resolved turbulent flows is  computationally very expensive.

Here we breifly summerize the algorithm put forward by Benettin \cite{benettin1980lyapunov} for  measuring multiple Lyapunov exponents. We begin by evolving a reference velocity field, $\boldsymbol{u_0}$, until it reaches a staistically steady state. We then make $M$ copies of this field, labelled $\boldsymbol{u_i}$, $i = 1 \dots M$. A unique small perturbation field is then applied to each copy. This perturbation field has a Gaussian distribution with zero mean and a variance of size $\delta_0$, chosen such that the perturbation may be considered infinitesimal. For each field we use the finite time Lyapunov exponent (FTLE) method \cite{ott2002chaos} and measure the growth of the difference fields $\boldsymbol{\delta_i}(t) = \boldsymbol{u_i} - \boldsymbol{u_0}$, rescaling the difference to its original size at time intervals of $\Delta t$ \begin{equation}
\boldsymbol{u_i}(\boldsymbol{k}, \Delta t) = \boldsymbol{u_0}(\boldsymbol{k}, \Delta t) + \frac{\boldsymbol{u_i}(\boldsymbol{k}, \Delta t) - \boldsymbol{u_0}(\boldsymbol{k}, \Delta t)}{\delta_0},
\end{equation} such that each perturbation continues to grow in the correct direction. The FTLEs are then given by \begin{equation}
\gamma_i(\Delta t) = \frac{1}{\Delta t}\ln \left(\frac{|\boldsymbol{\delta_i}(\Delta t)|}{\delta_0} \right),
\end{equation} and the Lyapunov exponents $\lambda_i$ are found by averaging over many iterations. Currently, this algorithm simply measures the largest Lyapunov exponent $M$ times, as this growth in this direction of the phase space will dominate all others. To circumvent this issue, we orthogonalize the $\boldsymbol{\delta_i}$ after each measurement of the $\gamma_i$ using the modifed Gram-Schmidt algorithm. For details of how this algorithm is defined for DNS of HIT see \cite{grappin1991lyapunov,deane}. If an infinite number of iterations were performed, this algorithm would return exponents ordered such that $\lambda_1 > \lambda_2 > \dots > \lambda_M$. As a result of a finite number of iterations, our spectra are not monotonically decreasing, however, the ordering achieved is reasonable as it is in \cite{keefe1992dimension, berera2019info}. This ordering property allows us to be confident we have found all positive exponents by choosing $M$ large enough that a tail of negative exponents persist after averaging. This orthogonalization step scales with $M^2$ and thus becomes a major bottleneck in these calculations. Additionally, we note that, as in \cite{keefe1992dimension,berera2019info}, we perform this procedure in the state space of the system, as opposed to in the tangent space as utilized in \cite{grappin1991lyapunov}. By ensuring our perturbation field is small enough, these two methods should give consistent results.
 
When using the stochastic forcing function, extra care must be taken in the implementation of this algorithm. If a new random force is generated for each of the $M$ copy fields, then the forcing acts as an effective perturbation every time-step and destroys the exponential divergence of the fields. Therefore, if a stochastic force is being used, the random force should be generated only once each time-step and then this force is applied to all fields.

\subsection{Sampling errors}\label{samp} 
In the computation of the Lyapunov exponents using the algorithm described in the previous section, an average must be performed to find the value of each exponent. The mathematical definition of the Lyapunov exponent calls for an average over an infinite number of iterations of the FTLE algorithm. Of course, in practice this cannot be done and only a finite number of iterations are performed. As a consequence, sampling errors are introduced into the computed mean value of each exponent. 

Such errors are further complicated in the case of turbulent fluid flow by the fact that, depending on the sampling frequency, the values obtained may be highly correlated. Typically, to avoid the complications these correlations cause, samples are taken at larger time intervals. However, given the massive numerical cost involved in the computation of many Lyapunov exponents, this is only viable for very low resolution cases.

\begin{figure}
    \includegraphics[width=\columnwidth]{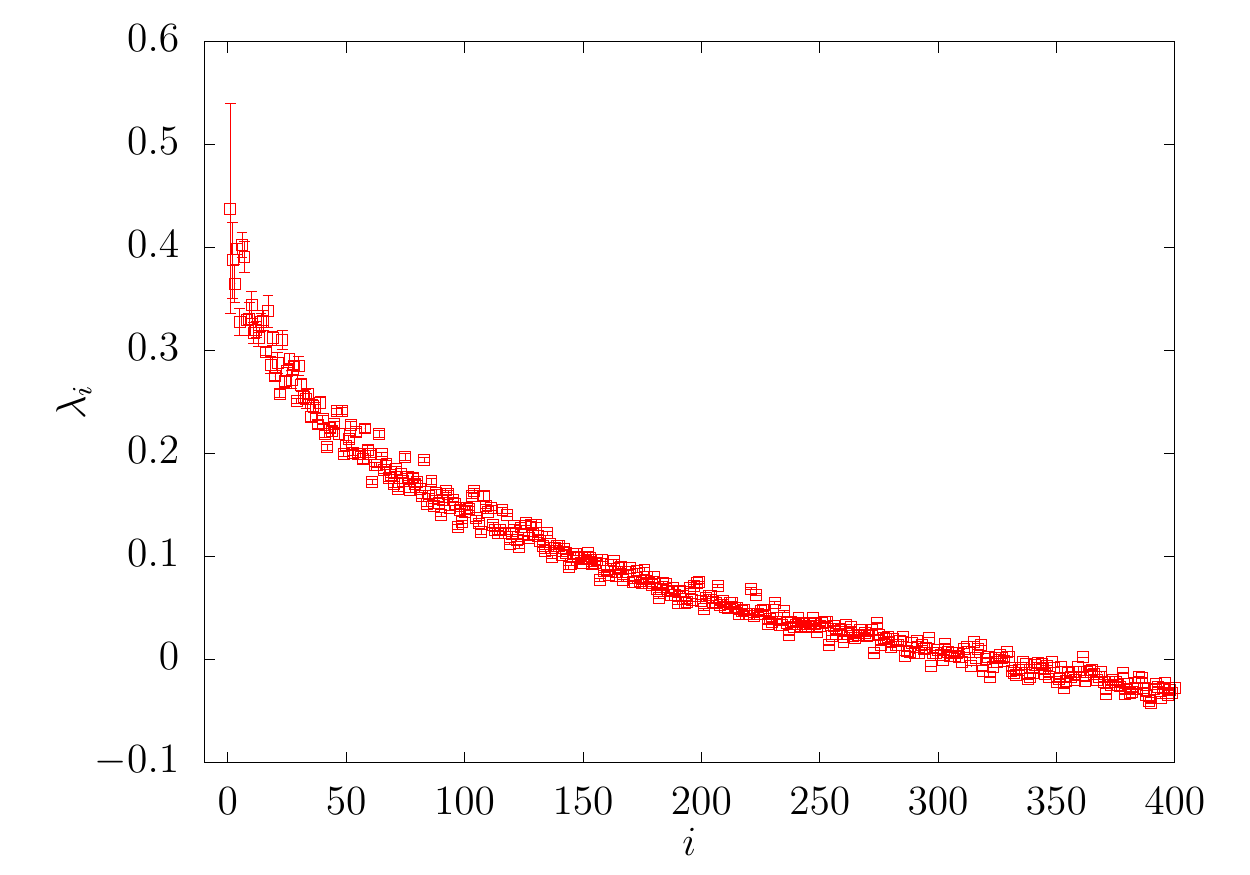}
    \caption{\label{erspec}Partial Lyapunov spectrum from $512^2$ simulation, highlighting that the error is concentrated in a small number of the largest exponents. }
\end{figure}

A number of methods for computing the sampling error of correlated data have been developed for use in DNS \cite{hoyas, oliver}. They share a common feature in that they both make use of an extension of the central limit theorem to weakly dependant variables, see \cite{billingsley} for details. We focus on the method detailed in \cite{oliver} and use it to find the standard deviation in our measurments of the exponents, and thus the Kolmogorov-Sinai entropy. Importantly, this method lets us make use of all the possibly correlated samples we have and as such is an efficient use of our computational effort. For the case of the entropy, these errors are tabulated in Tables \ref{tab:my-table}, \ref{tab:my-table1} and \ref{tab:my-table2}. Also listed are the number of samples taken for each simulation.

In Fig. \ref{erspec} we show a partial Lyapunov spectrum from a $512^2$ simulation. Here, it is clear that the largest exponents take the longest time to converge, as was reported in \cite{keefe1992dimension}. For the lesser exponents, their error bars are smaller than the points themselves. Given that the Kolmogorov-Sinai entropy is determined by the sum of a large number of exponents, the influence of the largest exponents reuduced convergence is damped by the quick convergence of the remaining exponents. As such, the largest relative errors are found in cases with very few positive exponents.

\section{Results}\label{4}

\subsection{Maximal Lyapunov exponent}\label{4a}

\begin{figure}
    \includegraphics[width=\columnwidth]{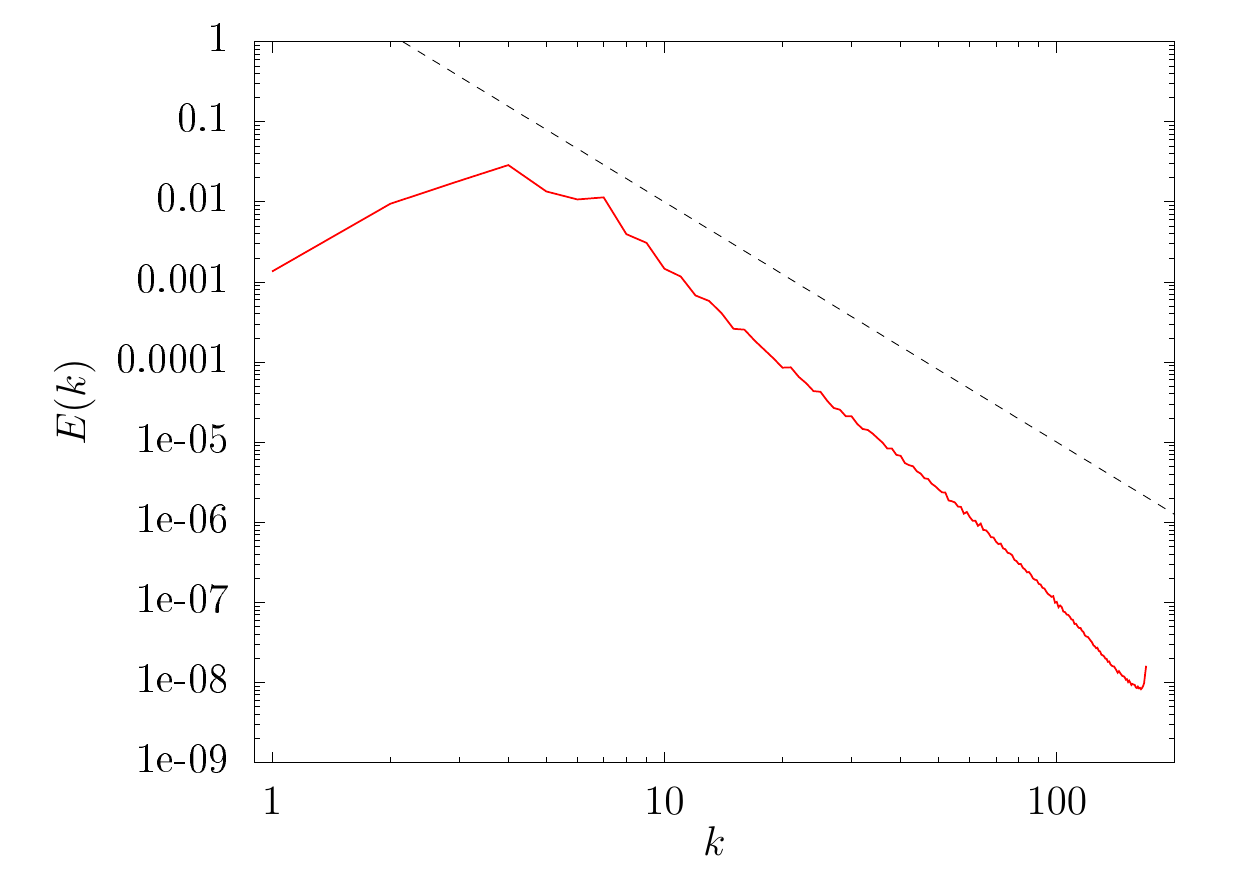}
    \caption{\label{espec}Energy spectrum from a $512^2$ simulation with $k_f = 7$ and $k_{\mathrm{min}} =  1$. Dashed line shows $k^{-3}$ scaling. }
\end{figure}

\begin{figure}
    \includegraphics[width=\columnwidth]{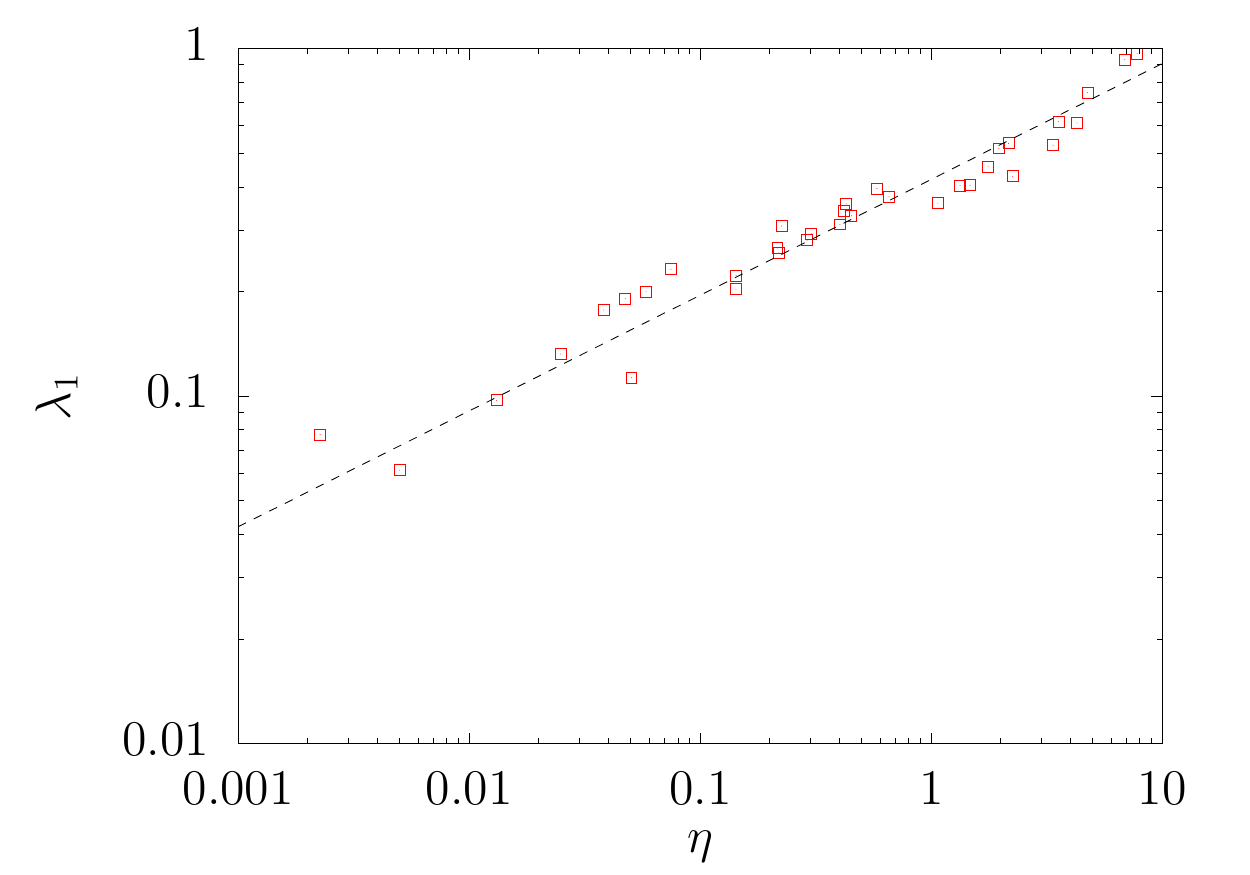}
    \caption{\label{fig1}Plot of the enstrophy dissipation rate, $\eta$, against the largest Lyapunov exponent,  $\lambda_1$. Dashed line shows the fit $0.42\eta^{1/3}$. }
\end{figure}

As discussed in section \ref{2}, depending on the form of the energy spectrum in the direct entrophy cascade region, there exist two possible scaling predictions for the maximal Lyapunov exponent, $\lambda_1$. The first of these predictions is valid if the energy spectrum takes the form $E(k) \sim k^{-3}$ and is given by Eq. \ref{eq1}, notably, this prediction has no Re dependence and is determined solely by the enstrophy dissipation rate, $\eta$. We note here that in our simulations, given the high computational demands imposed by the computation of a large number of Lyapunov exponents, we only achieve modest resolution. To illustrate this, in Fig. \ref{espec} we show the energy spectra from the highest resolution simulation in our data-set. It is clear that the spectrum in the enstrophy cascade region is steeper than $k^{-3}$. This is not surprising given the resolution achieved, and has been observed in previous studies \cite{legras,Ohkitani91}. In Fig. \ref{fig1} we show $\lambda_1$ against $\eta$ and we find our data is well fit by a power law of the form $\lambda_1 = \alpha \eta^{1/3}$, with $\alpha = 0.42 \pm 0.01$. The calculation of the maximal exponent only requires the simultaneous integration of two velocity fields and is much less resource intesive when compared to the calculation of the entropy and attractor dimension. As such, by computing the maximal exponent in separate simulations, the number of samples is far larger, $N_s \approx 5,000$ in all cases, leading to small error.
  
\begin{figure}
    \includegraphics[width=\columnwidth]{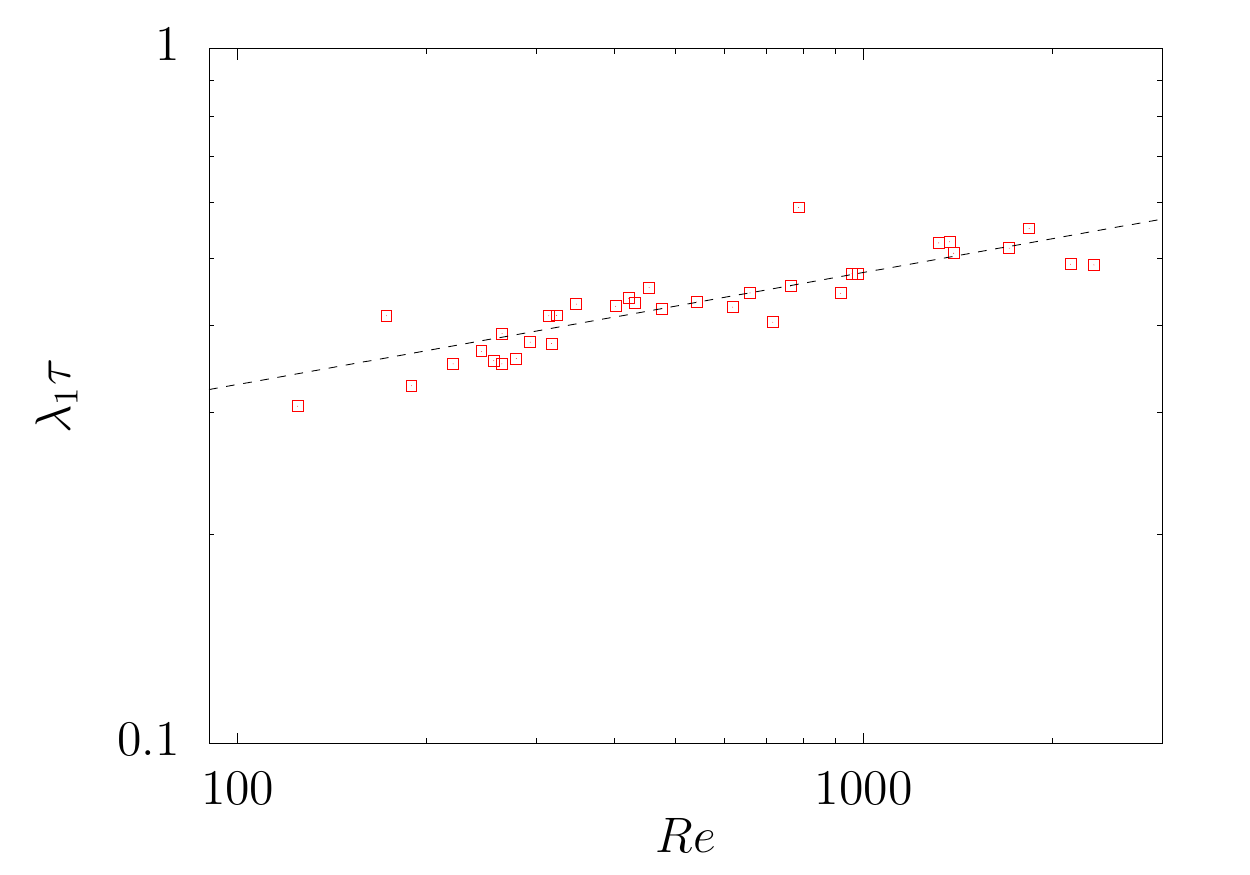}
    \caption{\label{fig2}Plot of Re against the largest Lyapunov exponent multiplied by the enstrophy dissipation time $\eta^{-1/3}$. Dashed line shows a power law fit with $0.16\mathrm{Re}^{0.16}$.}
\end{figure}

It was suggested by Kraichnan \cite{kraich71} that in order for the enstrophy flux to be constant in the direct cascade inertial range, then there should be a logarithmic correction to the energy spectrum. This alters the scaling prediction of Eq. \ref{eq1} to that of Eq. \ref{eq4} and introduces a dependence on Re. To test this we plot in Fig. \ref{fig2} the product $\lambda_1 \tau$, which, if Eq. \ref{eq1} is correct, should be constant for all Re, against the Reynolds number. In doing so, we find $\lambda_1 \tau$ slowly increases with $Re$, with the data being well fit by a power law of the form $\lambda_1 \tau = \beta \mathrm{Re}^{\gamma}$, where $\beta = 0.16 \pm 0.02$ and $\gamma = 0.16 \pm 0.02$. Due to the the difficulties in accurate measurement, we choose not to test the logarithmic scaling with $Re$ predicted in Eq. \ref{eq4}, however, Fig. \ref{fig2} does demonstrate that the maximal exponent does have a weak dependence on the Reynolds number. This weak dependence is in line with the logarithmic correction to the energy spectrum suggested by Kraichnan.

\subsection{Kolmogorov-Sinai entropy}\label{4b}

We turn now to the KS entropy for forced two dimensional turbulence. Once again, in Sec. \ref{2} we presented two scaling predictions derived via dimensional arguments. The first Eq. \ref{eq2} is derived for the case where there are no logarithmic corrections to the energy spectrum, whilst Eq. \ref{eq5} is valid with these corrections. Both cases have a dependence on the enstrophy dissipation time and the Reynolds number, with the corrected prediction introducing an additional logarithmic dependence on Re. Due to the cost of computing the KS entropy scaling quickly with Re (see \cite{berera2019info} for the three dimensional case which is more severe but illustrative), our results likely will not be able to quantify this logarithmic dependence, so we will focus on the prediction of Eq. \ref{eq1}.

Upon testing the prediction of Eq. \ref{eq2} against our data we find there is no scaling behavior and the value of $h_{\mathrm{KS}}\tau$ varies by orders of magnitude for the same value of Re. Within this data-set there are a range of different values used for the forcing length scale $k_f$, as well as three different physical box side lengths. If we fix the box side length at $2\pi$, we find what has the appearance of three separate close to parallel lines, one corresponding to each value of $k_f$. This shows there is some form of scaling with the integral scale Reynolds number, but that this is not the full picture. In \cite{grappin1987computation} the dimension of the attractor in two dimensional HIT was found to be dependent on $k_f$, however the exact dependence was not investigated. As such, the fact that our results for the entropy also show a $k_f$ dependence is not overly surprising, despite being at odds with Eq. \ref{eq2}. 

\begin{figure}
    \includegraphics[width=\columnwidth]{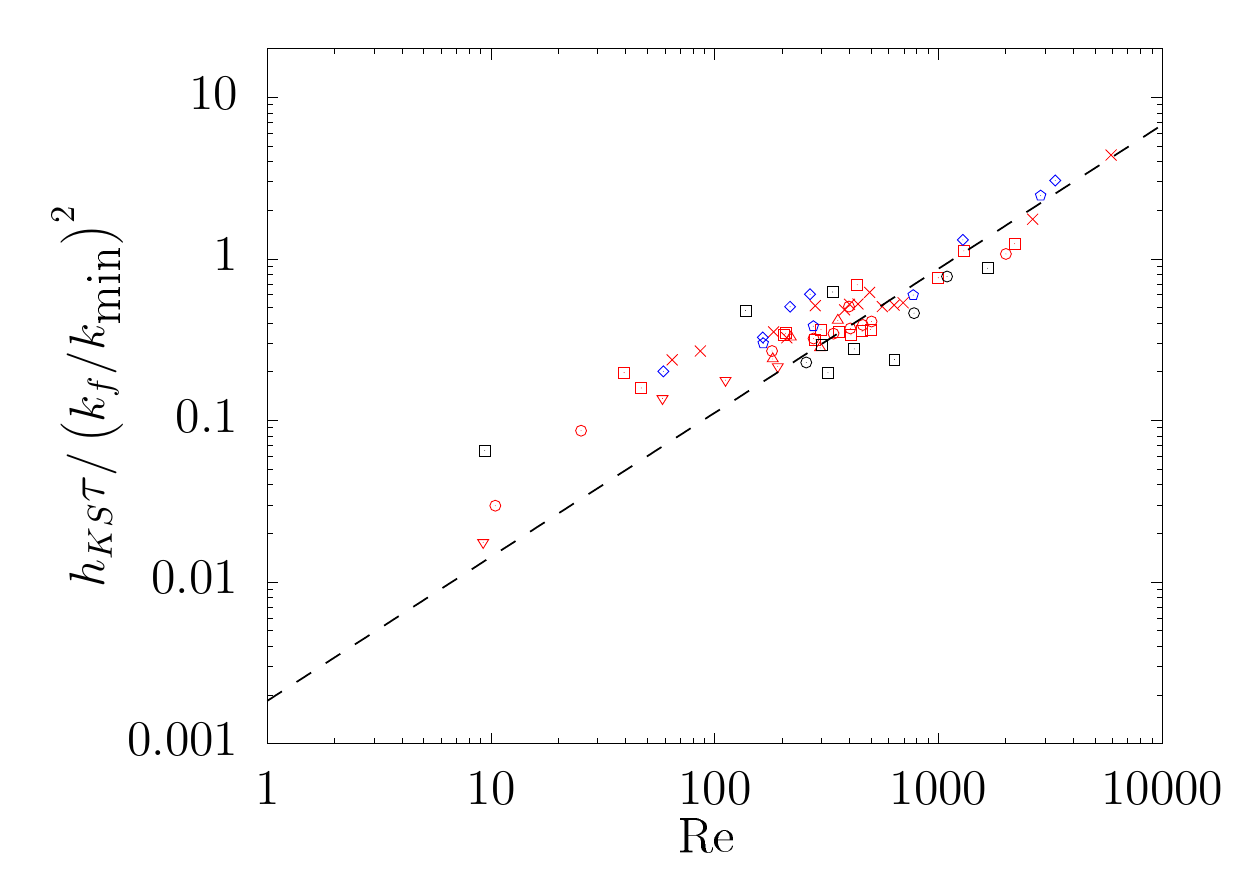}
    \caption{\label{fig5}Plot of Re against the Kolmogorov-Sinai entropy $h_{\mathrm{KS}}$ scaled by the enstrophy dissipation time-scale and the ratio of $k_{f}$ to $k_{\mathrm{min}}$, fit corresponds to 0.0018Re$^{0.9}$. The values of ($k_{\mathrm{min}}$,$k_f$)  are varied and displayed as $(1,3)$ red ($\times$), $(1,5)$ red ($\square$), $(1,7)$  red ($\circ$), $(1,9)$ red ($\bigtriangleup$), $(1,11)$ red ($\bigtriangledown$), $(2,4)$ blue ($\Diamond$), $(2,6)$ blue ($\pentagon$), $(4,8)$ black ($\square$) and $(4,12)$ black ($\circ$).}
\end{figure}

In order to correct the prediction in Eq. \ref{eq2} to account for this forcing scale dependence, we will consider what was found in \cite{grappin1991lyapunov}, where it was observed that the attractor dimension grew at the same rate as the number of modes in the inverse energy cascade inertial range. Using this a reasonable ansatz for the correction factor, $C$, is \begin{equation}\label{fact}
C \sim \left(\frac{k_f}{k_{\mathrm{min}}}\right)^2,
\end{equation} where $k_{\mathrm{min}}$ is determined by the side length, $x$, of the box our fluid resides within using \begin{equation}
k_{\mathrm{min}} = \frac{2 \pi}{x}.
\end{equation} The typical choice in simulations is $x=2\pi$ restricting the allowed wavenumbers to integer values. By choosing $x=\pi$ and $x=\pi/2$ we have $k_{\mathrm{min}} = 2$ and $k_{\mathrm{min}} = 4$ respectively. Since energy is injected at $k_f$, then a natural lower bound for the inverse cascade is $k_{\mathrm{min}}$, and thus $C$ has the desired scaling behavior. We then consider a corrected scaling prediction for the KS entropy of the form \begin{equation}
h_{\mathrm{KS}} \tau \left (\frac{k_{\mathrm{min}}}{k_f}\right)^2 \sim \mathrm{Re} .
\end{equation} Using this new scaling prediction, our data is shown in Fig. \ref{fig5}. It can be seen that all points fall on a straight line, thus indicating a power law scaling. We find the data is well fit by a power law of the form $h_{\mathrm{KS}} \tau \left ({k_{\mathrm{min}}}/k_f\right)^2 = a\mathrm{Re}^{b}$, with $a = 0.0018 \pm 0.0005$ and $b = 0.9 \pm 0.03$. There is a notable spread in this data which we do not believe to be solely the result of sampling errors, which are reasonably small in general. Instead, this spread is likely explained by the use of the correction factor $C$ defined in Eq. \ref{fact}. This factor does not contain any information regarding the structure of the underlying flow and is merely a ratio of length scales. However, without the use of such a term, no scaling at all is found for simulations with differing $k_f$ and $k_{min}$ values. It is likely a more flow specific correction can be found, but we do not pursue that here. 

\begin{figure}
    \includegraphics[width=\columnwidth]{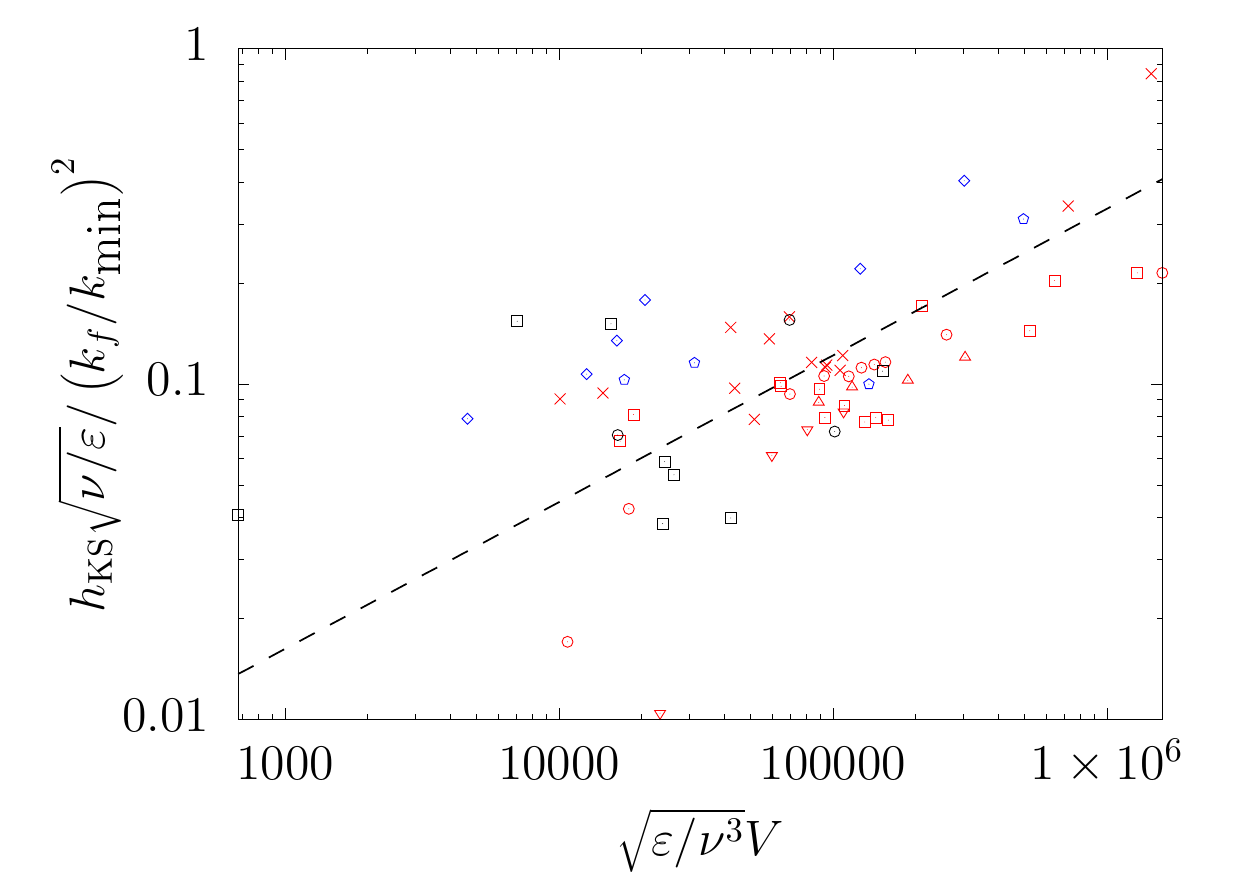}
    \caption{\label{fig5a} Plot of the scaling prediction for the Kolmogorov-Sinai entropy given by Ruelle and Lieb. The values of ($k_{\mathrm{min}}$,$k_f$)  are varied and displayed as $(1,3)$ red ($\times$), $(1,5)$ red ($\square$), $(1,7)$  red ($\circ$), $(1,9)$ red ($\bigtriangleup$), $(1,11)$ red ($\bigtriangledown$), $(2,4)$ blue ($\Diamond$), $(2,6)$ blue ($\pentagon$), $(4,8)$ black ($\square$) and $(4,12)$ black ($\circ$).}
\end{figure}

We now turn our attention to the scaling prediction given in Eq. \ref{eq7}. It is interesting that this prediction must also be corrected by $C$, or else the same issue of different scaling behaviors for each value of $k_f$ and $k_{\mathrm{min}}$ appears once more. We show this in Fig. \ref{fig5a} in which we have non-dimensionalized the entropy using $\sqrt{\nu/\varepsilon}$. The spread in the data here is more pronounced than for the simpler scaling of Eq. \ref{3}; once more this is a combination of small sampling errors and the use of the correction factor $C$, the effect may be exacerbated in this case by the logarithmic scale and small values on the y-axis.

Our data thus shows that the scaling of the Kolmogorov-Sinai entropy for two dimensional turbulence exhibits a dependence on both the forcing length scale and the system size. This is very much at odds with the picture in three dimensional turbulence, where we found the the scaling of the entropy is very close to that predicted in the K41 theory, depending only on small scale features of the flow \cite{berera2019info}. This is perhaps best explained by considering the nature of the triadic interactions in two dimensional turbulence. In \cite{triad}, it was shown that non-local triad interactions have an important effect on both the energy and enstrophy inertial ranges. When viewed in physical space this manifests itself in the appearance of long-lived coherent vortices, which then influence the small scales. As such, the fact that in two dimensions the large scales have a direct effect on the chaotic properties of the flow should not come as a surprise.

\subsection{Attractor dimension}\label{4c}

By computing a large enough subset of the Lyapunov spectrum, it is also possible to make an estimate of the dimension of the attractor for forced two dimensional HIT. To do so, we make use of the Kaplan-Yorke conjecture \cite{kaplan1979chaotic}, which suggests the attractor dimension can be found using \begin{equation}
\mathrm{dim}(A) = j + \frac{\sum_{i=0}^{j} \lambda_i}{|\lambda_{j+1}|},
\end{equation} in which $j$ is the index of the Lyapunov exponent such that\begin{equation}
\sum_{i=0}^{j} \lambda_i \geq 0, \quad \mbox{and} \quad \sum_{i=0}^{j+1} \lambda_i < 0.
\end{equation} From this definition, it is clear that the computation of the attractor dimension will require more exponents than needed for the Kolmogorov-Sinai entropy. This definition makes quantifying the effect of the error in the Lyapunov exponents on the attractor dimension difficult. Any fluctuation in values of the exponents will effect the value of $j$ in a complex manner. As such, we do not include the standard deviation of the attractor dimension in either Table \ref{tab:my-table}, \ref{tab:my-table1} or \ref{tab:my-table2}, but it is likely to be of the order of the error in the Kolmogorov-Sinai entropy as both quantities are derived from the same data.

\begin{figure}
    \includegraphics[width=\columnwidth]{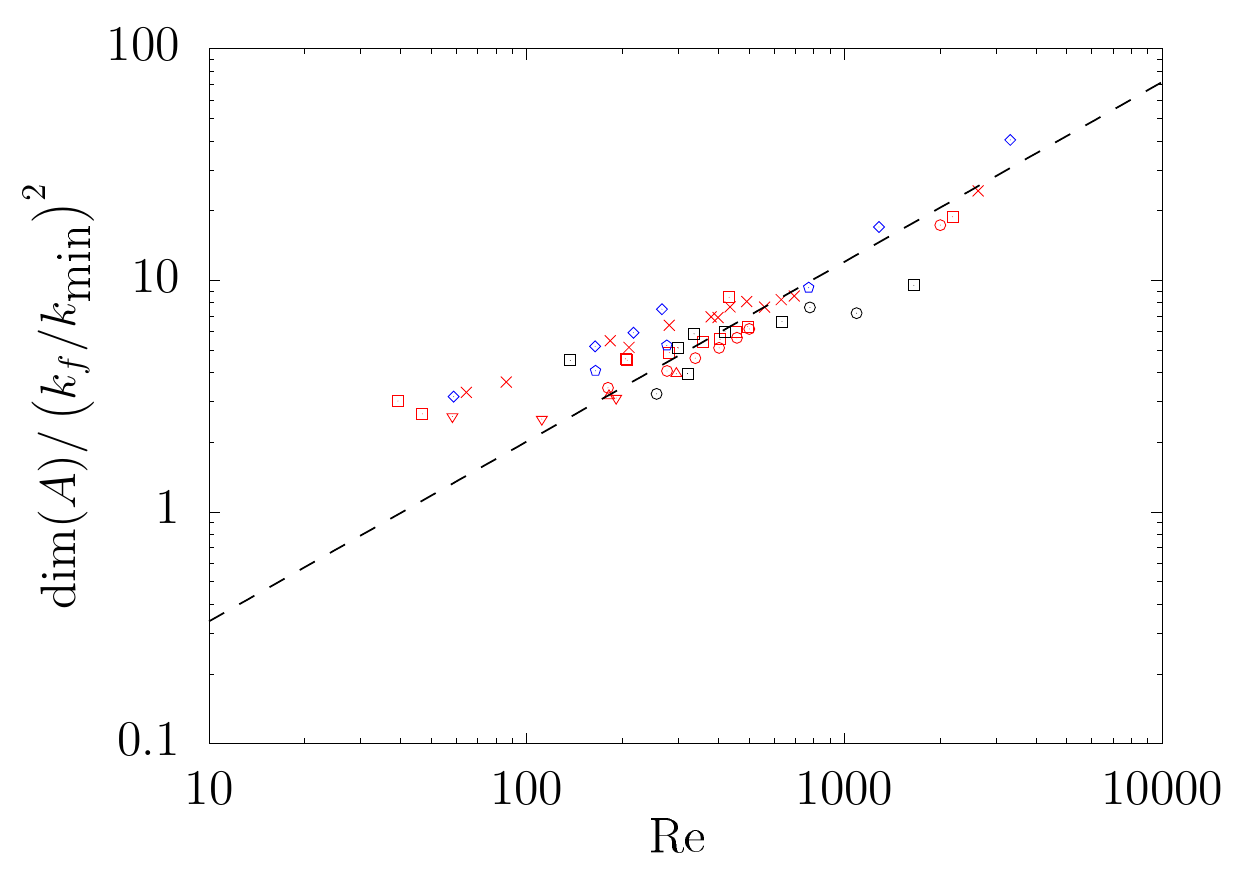}
    \caption{\label{fig6}Plot of Re against the attractor dimension, $\mathrm{dim}(A)$ scaled by the ratio of $k_{f}$ to $k_{\mathrm{min}}$, fit corresponds to 0.055Re$^{0.78}$. The values of ($k_{\mathrm{min}}$,$k_f$)  are varied and displayed as $(1,3)$ red ($\times$), $(1,5)$ red ($\square$), $(1,7)$  red ($\circ$), $(1,9)$ red ($\bigtriangleup$), $(1,11)$ red ($\bigtriangledown$), $(2,4)$ blue ($\Diamond$), $(2,6)$ blue ($\pentagon$), $(4,8)$ black ($\square$) and $(4,12)$ black ($\circ$).} 
\end{figure}

\begin{figure}
    \includegraphics[width=\columnwidth]{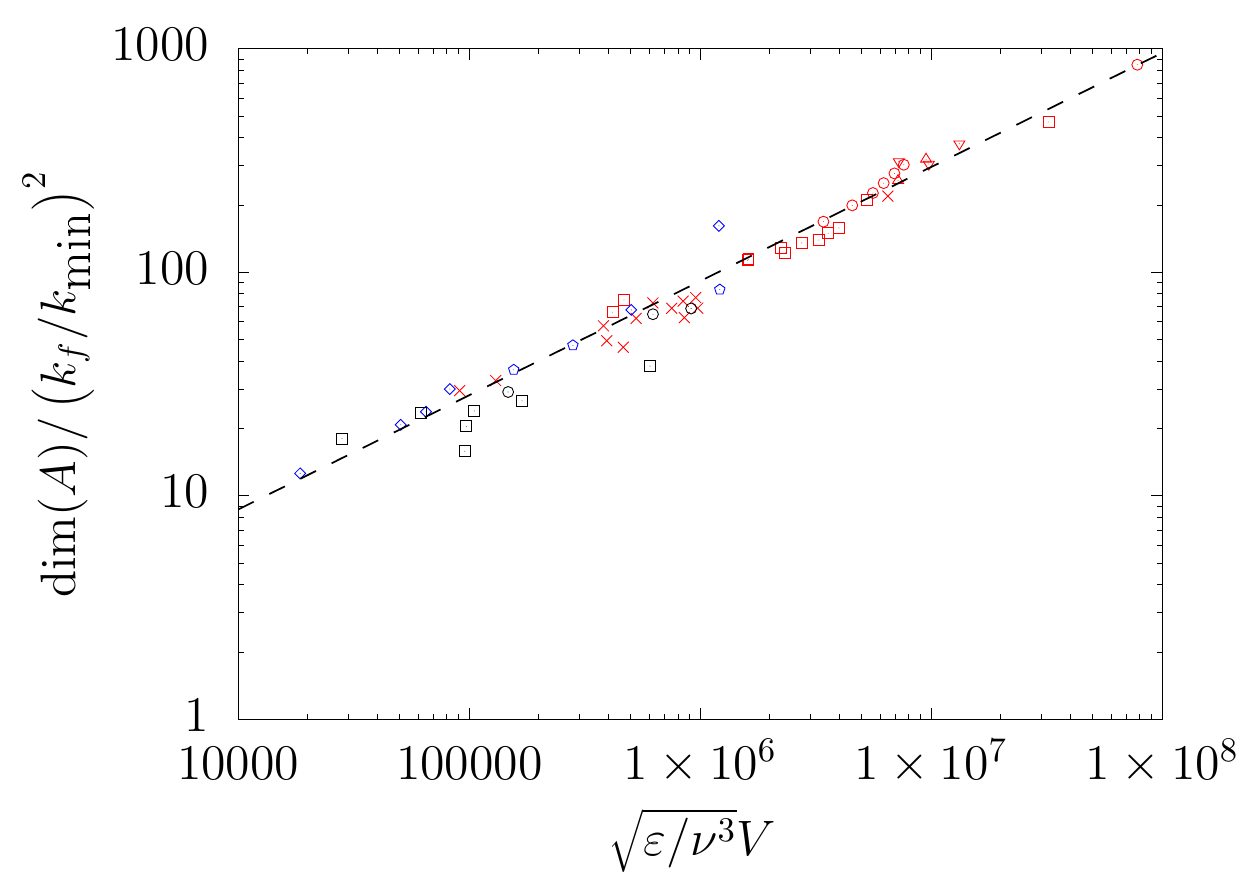}
    \caption{\label{fig6a}Plot of the attractor dimension scaling prediction given by Ruelle and Lieb. The values of ($k_{\mathrm{min}}$,$k_f$)  are varied and displayed as $(1,3)$ red ($\times$), $(1,5)$ red ($\square$), $(1,7)$  red ($\circ$), $(1,9)$ red ($\bigtriangleup$), $(1,11)$ red ($\bigtriangledown$), $(2,4)$ blue ($\Diamond$), $(2,6)$ blue ($\pentagon$), $(4,8)$ black ($\square$) and $(4,12)$ black ($\circ$).}. 
\end{figure}

Naturally, the numerical computation of the attractor dimension comes at a severe computational cost. However, as with the entropy, when compared to the three dimensional case, the calculation for the attractor dimension is more favorable and a reasonable measurement of the scaling behavior can be made. The results of this measurement are displayed in Fig. \ref{fig6} in which we have plotted against Re the scaling prediction of Eq. \ref{eq2} corrected by the scaling factor $C$ described previously. Upon doing so, we find the data is well fit by a power law of the form $\mathrm{dim}(A)\left ({k_{\mathrm{min}}}/k_f\right)^2 = c\mathrm{Re}^{d}$ with $c = 0.055 \pm 0.02$ and $d = 0.78 \pm 0.04$. As with the entropy, we also find that when considering the Ruelle-Lieb prediction of Eq. \ref{eq8}, the correction factor is once again necessary and this is demonstrated in Fig. \ref{fig6a}. Although the scatter is less in these figures, it is still present. This is again a result of a combination of sampling error and the use of the corrective term $C$. Notably, it is clear that data points with either low Re or higher $k_{\mathrm{min}}$, which have the fewest postive exponents, show the largest spead.

The attractor dimension gives a measure of the total number of active degrees of freedom in the flow. In \cite{grappin1991lyapunov}, the attractor dimension was found to grow with the width of the energy inertial range. Our results are in agreement with this finding, although we also find a contribution from the enstrophy inertial range due to the dependence on the ratio of large to small scales in the flow measured by Re.

\subsection{Lyapunov spectrum}

It is also of interest to investigate the shape of the Lyapunov spectrum, in particular, the distribution of exponents about $\lambda \approx 0$. It was suggested by Ruelle \cite{ruelle1982large, ruelle1983five} that it may be possible for the distribution of exponents to become singular about this point. Using the GOY shell model \cite{yamada1987lyapunov, yamada1988lyapunov} it was found that in both two and three dimensions the distribution of exponents did indeed become singular. However, it was later suggested that this divergence was caused by the numerical discretization employed in these works \cite{yamada1998asymptotic}.

\begin{figure}
    \includegraphics[width=\columnwidth]{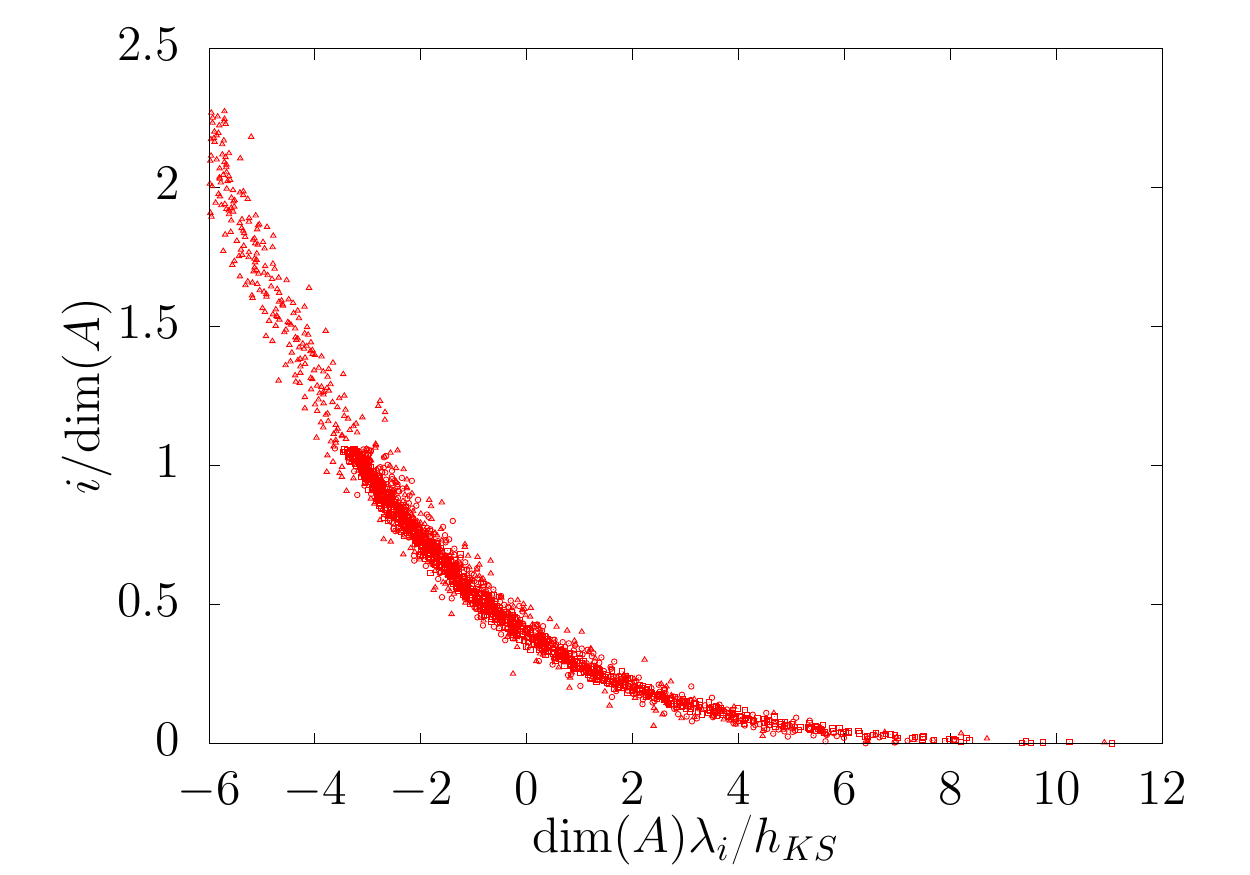}
    \caption{\label{fig7} Lyapunov spectra normalized by $h_{KS}$ and dim$(A)$. The results of a number of simulations are shown here to highlight the similarity property of the spectra.} 
\end{figure} 

In Fig. \ref{fig7}, we show the Lyapunov spectra from a number of our simulations scaled by both their Kolmogorov-Sinai entropies and the attractor dimensions, such that the spectra collapse onto a single curve. From this figure, it is clear there is no divergence around $\lambda \approx 0$ in our simulations. This is consistent with what was found in three dimensional turbulence \cite{keefe1992dimension,berera2019info,spec}, although it should be noted that in \cite{spec} a `knee'-like structure was found around $\lambda \approx 0$, which is also seen in  simulations of Rayleigh-B{\'e}nard convection \cite{xu}. This structure does not, however, appear to be a true divergence.

\section{Conclusion}\label{5}

This work has been focused on the calculation of a number of standard measures of chaos in two dimensional forced incompressible HIT using pseudospectral DNS. A number of scaling predictions for these quantities have been made in the literature and, using our numerical results, we have tested a subset of them. It was found that the maximal exponent displays a weak dependence on the Reynolds number of the flow, consistent with the logarithmic correction to the energy spectrum suggested by Kraichnan. However, it was seen that for the Kolmogorov-Sinai entropy and attractor dimension, the predictions made on dimensional arguments were not sufficient. Corrections relating to the forcing length scale and system size were then found to be required. It is suggested that these corrections are required due to non-local effects in two dimensional turbulence as a result of coherent vortices. 

When comparing these results to three dimensional turbulence, it is found that these chaotic properties scale with Re far more slowly in two dimensional turbulence. Futhermore, since these chaotic properties depend on large scale details of the flow in two dimensions, as opposed to only on small scale features in three dimensions, they provide further evidence of non-universality \cite{danilov} in two dimensional turbulence. Given the two dimensional phenomenology seen in the atmospheres of the Earth and Jupiter \cite{nastrom,young}, this may have important implications for atmospheric predictability. In reality, this two dimensional phenomenology is not the entire story, as such systems are likely more accurately described by thin layer turbulence \cite{benav,split}. In thin layer turbulence, it is found that there are a number of critical points where the system transitions from purely three dimensional behavior to coexisting two and three dimensional phenomenology, and then from this state to purely two dimensional turbulence \cite{benav}. It is in fact not just thin layer systems in which this kind of behavior is seen. Indeed, in systems undergoing rotation, as well as stratification and influence from an external magnetic field, a similar transition from three dimensional to two dimensional behavior is seen \cite{smith, sozza, alexak}. The predictability of such would be of interest to study and compare with the idealized cases of pure HIT in two and three dimensions.

Such is the complexity of atmospheric systems, that even all of the variants discussed above only begin to scratch the surface. As such, simplified models which approximate the true dynamics of the atmosphere are often used. These models also exhibit sensitivity to initial conditions and, in some cases, Lyapunov spectra have been measured \cite{atmos}. In one such case in a coupled atmosphere-ocean model \cite{atmos1}, a large number of near zero Lyapunov exponents are found, suggesting a possible divergence in the spectra. Whether this divergence is simply a feature of the simplified model or of the true dynamics is an interesting question, however, given the computation cost of even the simple case studied in this work, its answer is likely some way off.

\begin{acknowledgments}
This work used the Cirrus UK National Tier-2 HPC Service at EPCC \cite{cirrus} funded
by the University of Edinburgh and EPSRC (EP/P020267/1).
D.C. is supported by the University of Edinburgh.
A.B. acknowledges funding from the U.K. Science and Technology Facilities Council.
\end{acknowledgments}

\end{document}